# Ion Implantation for Deterministic Single Atom Devices

J. L. Pacheco, M. Singh, D. L. Perry, J. R. Wendt, G. Ten Eyck, R. P. Manginell, T. Pluym, D. R. Luhman, M. P. Lilly, M. S. Carroll and E. Bielejec.
Sandia National Laboratories, Albuquerque, New Mexico 87185, USA

**Abstract:**
We demonstrate a capability of deterministic doping at the single atom level using a combination of direct write focused ion beam and solid-state ion detectors. The focused ion beam system can position a single ion to within 35 nm of a targeted location and the detection system is sensitive to single low energy heavy ions. This platform can be used to deterministically fabricate single atom devices in materials where the nanostructure and ion detectors can be integrated, including donor-based qubits in Si and color centers in diamond.

Deterministic placement of single atoms is a key capability for fabrication of nanometer scale and single atom solid-state devices in a range of material systems including Si, diamond, and III-V compounds. Examples of Si-based devices include: donors coupled to quantum dots [1] for charge [2], electron [3, 4], and nuclear spin [5, 6] qubits (quantum bits) and acceptors coupled to silicon cavities to create phononic qubits [7]. Single color (defect) centers in diamond have a range of applications including metrology [8], quantum computing using nitrogen-vacancy (NV) centers [9] and coupling silicon-vacancy (SiV) centers to photonic cavities for cavity QED experiments [9, 10]. In III-V materials, deterministic seeding of nucleation sites for controlling the quantum dot growth locations [11] has many potential applications including the development of single photon sources [12]. In many applications, placement of single ions within small volumes is critical. Ion implantation has been widely applied in the semiconductor industry for introducing dopants with a nominal depth and dose by varying the implant energy and the exposure time, respectively. The key challenges to extending this technique down to single atom control are the precise control over the atom's position and the implantation of one and only one atom. Techniques include in-situ ion detection using PIN diode detectors [13-15] and FinFETs [16, 17] and detection of secondary electrons [18].

We present a "top down" ion implantation approach to deterministic single atom device fabrication in Si and in other material systems suitable for ion detection including diamond [19] and GaAs [20]. This requires the ability to place the implanted ions with high positioning precision and deterministic control over the number of ions implanted. We use the nanoImplanter (nI) at Sandia National Labs (SNL), which is a direct-write focused ion beam platform to control the positioning of the implanted ion and *in-situ* solid-state detectors for single ion detection. We demonstrate single ion targeting to less than 35 nm allowing for deterministic single ion implantation. The combination of focused ion beams, direct write lithography, fast beam blanking and chopping, ion mass selectivity, in-situ detection and electrical probing are key features that enable rapid prototyping, customized implantation and high throughput fabrication of deterministic single atom devices. As a test of our "top-down" ion implantation and detection capability we have used counted ion implantation to successfully create donor-based qubit prototype devices [21].

Focused ion implantation is performed at the Ion Beam Laboratory (IBL) at SNL using the nI. The nI is a 100 kV focused ion beam (A&D FIB100nI) with a three-lens system designed for high mass resolution using a Wien, or ExB, filter and single ion implantation using fast beam blanking. Ion beams from a variety of liquid metal alloy ions sources (LMAIS), such as AuSiSb, can be extracted. In this manner, the nI allows for high-brightness ion beams from ~1/3 of the periodic table. Figure 1 shows a schematic diagram of the nI. The first aperture controls the beam current. The extractor and first condenser lens voltage focus the ion beam at the ExB filter. The ExB and second aperture define the mass resolution; we have demonstrated a mass resolution of (M/ΔM) > 61 sufficient to separate the individual isotopes of Sb and Si for example. This is a key consideration for single atom devices where the nuclear spin properties may govern device operation [22]. The fast beam-blanker voltage rise-time and the size of the chopping aperture set the minimum ion beam gating pulse length to be ~16 ns. Throughout the paper, this is referred to as pulse or ion beam pulse. This capability gives explicit control over the average number of ions implanted. The third aperture and the objective lens create the third crossover point at the target plane and define the beam spot size, typically ranging from 10 to 50 nm. The nI is a direct write lithography platform combining a high resolution laser interferometry driven stage and electrostatic draw deflectors to position the beam with a step size of 0.8 nm when using a 50 μm write field (Raith Elphy Plus). Single ion positioning precision is determined by the beam spot size on target convolved with the longitudinal and lateral ion distribution profile [23], or straggle, in the substrate.

For single ion detection, we use SNL fabricated avalanche photo-diodes (APD) detectors [24]. Figure 2a shows an optical image of an APD with the construction zone, where a nanostructure can be fabricated (Fig. 2b), within the active detection region (between the anode and the cathode). Figure 2c shows a SEM image of an accumulation-mode nanostructure, the light grey areas are the poly-Si gates and darker regions are where the poly-Si has been removed. The pattern contains the necessary electrodes (right plunger (RP), center plunger (CP), left plunger (LP)) to electrostatically define a single electron transistor below the accumulation gate (AG) [25] and perform quantum transport measurements by monitoring the source-drain current ($I_{SD}$). Figure 2d shows a cross-sectional view of the detector. The central $P^+$ region corresponds to 50 keV and 140 keV B implants each at a dose of $4\times10^{12}$ cm$^{-2}$. The $N^+$ cathode has an As implant at 30 keV at a dose of $3\times10^{15}$ cm$^{-2}$, the guard ring has a P implant at 75 keV at a dose of $4\times10^{11}$ cm$^{-2}$ and the P+ anode has a BF$_2$ implant at 90 keV at a dose of $1.5\times10^{15}$ cm$^{-2}$. An electric field, intrinsic or applied, is established in the active detection region to drive ion detection.

These detectors can be operated in Geiger mode (GM) [26] or linear mode (LM). In LM, the detector is operated in the low reverse bias regime or zero bias with only the intrinsic fields present. The detector signal is proportional to the energy deposited as ionization, providing a methodology to determine the number of ions implanted for a given energy. In this paper, we employ detectors operated in LM at zero bias.

Figure 2e shows a diagram of the LM detection setup. Electron-hole pairs generated in the Si substrate during an ion strike induce a charge transient that is measured, which we call detector signal or detector response throughout this paper. This method of detection is known as ion beam induced charge collection, or IBIC [13, 27]. The detector signal is first amplified by an Amptek A250 CoolFET and filtered using an Stanford Research Systems (SRS) SIM965 to remove high frequency noise (> 6kHz). Two SRS box-car modules (SR250), referenced to the ion beam pulse, sample the filtered detector signal before and after an ion strike. The two box-car signals are then subtracted using an SIM980 module resulting in a measurement of the detector signal amplitude due to the ion strike. For the experiments described here, the nI is operated with an accelerating voltage of 100 kV and a Si++ beam with energy of 200 keV is selected using the ExB filter. This beam spot size typically is < 25 nm. To set the average number of ions per ion beam pulse, we measure the beam current on a Faraday cup and then adjust the beam pulse length to deliver the desired average number of ions. For the dataset in Fig 3a we measured a beam current of 0.56 pA, and set the pulse length to 1µs pulse to achieve an average of 1.75 ions on target per pulse.

The intrinsic detection efficiency can be written as the following [28]:

$$\epsilon_{int} = \frac{N_r}{N_i},$$

where $N_r$ is the number of detected ions (detector response) and $N_i$ is the number of incident ions. For a pulsed ion beam, the above expression becomes a sum over all ion beam pulses and detector responses recorded but, for a constant beam current and pulse width, the summation symbols are redundant and omitted here for simplicity.

$N_i$ is determined by measuring the beam current into the Faraday cup and setting the beam pulse width to produce the desired average number of ions per pulse (as described above). The number of ions incident will be $N_i = \frac{I_{beam}}{q} \tau N_p$, where $I_{beam}$ is the ion beam current measured, $q$ is the ion charge state, $\tau$ is the ion beam pulse width, and $N_p$ is the number of ion beam pulses. Equivalently, since the number of ions per ion beam pulse follow a Poisson distribution; we can write $N_i$ as:

$$N_i = N_p \sum_{k=0}^{infinity} k \frac{e^{-\mu} \mu^k}{k!} = \sum_{k=0}^{infinity} k\, P(k),$$

where $k$ is the Poisson variable (number of ions in a pulse) and $\mu$ is the average success rate (average number of ions in a pulse) based on the beam current and the ion beam pulse width. Note that the above expression holds because the sum of two independent Poisson variables is a Poisson variable itself.

Similarly, $N_r$ can be written as:

$$N_r = N_p \sum_{k=0}^{infinity} k\, A(k),$$

where $A_k$ is the normalized area under the curve for the peak corresponding to $k$ number of ions, and $N_p$ is the number of pulses (including those with zero ions).

We experimentally determined $N_r$ from the IBIC response spectrum of the detector as shown in Fig. 3a, for 200 keV Si ions. We exposed a construction zone where the poly-Si was removed leaving only a uniform 7 nm gate oxide to an average of 1.75 ions per pulse. The detector response amplitude was recorded for each implantation event to generate the spectrum shown. The quantization of the number of ions per pulse is represented by the peaks in the spectrum, which correspond to 0, 1, 2, 3, and 4 ions per pulse. Each of the peaks was fitted with a Gaussian [28]. The overall fit to the data (dashed line) has an adjusted $R^2$ of 0.975 [29]. The area under the overall fit was used to normalize the area under each individual peak and the resulting ratio of the area under each peak to the total area corresponds to the probability, $A(k)$, that 0, 1, 2, 3 and 4 ions were implanted, respectively. Note - we truncate the series expressions for $N_r$ and $N_i$ at $k = 4$ because we could no-longer accurately fit the higher ion peaks (> 4 ions) in the experimental dataset. From this, the detection efficiency is calculated to be ~100%. We point out that the normalized detector signal amplitudes agree to ± 3% of the probability predicted by Poisson statistics. Errors not accounted for in this calculation are primarily due to peak fitting in Fig. 3a, fluctuations in beam current or beam current measurement (up to ± 0.02pA). The centroids of the Gaussian fits are plotted in Fig. 3b as a function of the number of ions, where the error bars are ± σ from each of the corresponding Gaussian fits. This plot can be treated as a calibration curve where the magnitude of the detector response can be converted to the number of ions implanted. The detector response is linear in the few- ion per pulse regime with ~0.037 V per 200 keV Si ion implanted.

Deterministic single ion implantation is accomplished by setting the ion beam pulse to deliver, on average, a fraction of an ion per pulse. Figure 3c shows the detector response spectrum for an average of 0.1 ions per pulse. Fitting the 0 and 1 peaks with Gaussian functions shows that ~95% of the events show no detector response and ~5% show a detector response that corresponds to a single ion; a single ion implant. The probability of 2 or more ions per pulse is < 0.05% from Poisson statistics. A signal-to-noise ratio (SNR = $\mu_s/(\mu_n+\sigma_n)$) of ~21 is calculated from the average single ion response ($\mu_s = 0.037$) and one standard deviation of the noise ($\sigma_n = 0.0017$) (the centroid of the noise peak, $\mu_n = 0$). The expected error rate as determined by the overlap between the 0 (noise) and 1 ion (signal) peaks is negligible in this case, see Table 1. This result demonstrates the ability to perform deterministic single ion implantation.

To determine the spatial accuracy of implantation, we performed high-resolution IBIC scans on a nanostructure patterned within the construction zone. The Raith patterning software is used to align the ion beam to the lithography pattern using ion beam generated- secondary electron images of the alignment marks. A 100 by 100 implant array with a 20 nm pitch centered on the nanostructure was defined in the patterning software. The beam pulse length was set to deliver an average of 2 ions per pulse at each implant location.

Figure 3d shows an overlay of nanostructure CAD and the X, Y positions where a detector response was measured above a threshold of 0.01 V (set to detect a > 1 ion response). The sparse red dots that appear over poly-Si gates regions (see Fig. 2c) are likely due to implant events where 200 keV Si ions go through the poly-Si stack (average range of ~65 nm below the $SiO_2$/Si interface) and generate a signal above the threshold. We numerically optimized the alignment between the detector response map and the nanostructure CAD drawing by shifting the detector response map. We find that the center of the detector response map and the CAD drawing are offset by 35 ± 2 nm in X and 10 ± 2 nm in Y for the data set shown. Analysis of additional data sets show offsets with approximately the same magnitude or less, indicating a spatial targeting resolution of < 35 nm. This is a ~2x improvement on state-of-the-art [30-32]. Importantly, Fig. 3d also illustrates the successful integration of ion detection and nanostructure fabrication.

Our "top down" single ion implantation technique is compatible with standard CMOS processes and presents a path forward to deterministic fabrication of single atom devices. The results presented thus far are for 200 keV Si. However, we have used this same technique to fabricate operational qubit prototype devices with a counted number of 120 keV Sb donors (implanted through a 35 nm gate oxide, targeted implant location is shown in Fig. 2c by red dot) [21]. Table 1 shows the range and straggle (calculated using Stopping and Ion Range in Matter, SRIM [23]), signal-to-noise ratio (SNR), and detection error rates for several of the ion/energy combinations tested. Lowering the implantation energy results in an improved positioning precision by reducing the straggle. As currently configured we have detected single Sb ions down to 20 keV through a 7 nm gate oxide with a SNR of 2.5. The detection efficiency is 87.6% estimated using the same calculation as outlined above resulting in a combined error rate of detection of ~20%. 20 keV Sb implantation into a Si substrate corresponds to ~1400 e-h pairs per ion (SRIM). Experimental results suggest that this number may be lower by a factor of ~3 [33, 34]. Notwithstanding, high fidelity detection of < 1000 e-h pairs is a challenge. By combining our targeting implantation capabilities with 10 nm PMMA mask openings and lowering the implantation energy, the resulting donor position can be localized to a sphere of ~10 nm in radius centered at the end of range. Recent device simulations suggest that this positioning precision is sufficient for the fabrication of donor-donor (2-qubit) [35], donor chains [36] for CTAP (coherent tunneling adiabatic passage) processes [37], and electron shuttling devices [38].

If we consider a high-density array of deterministic single atom sites on an active detection substrate, then what is the array pitch, time per implant, and size of the implantation area that this 'top-down' technique can provide? Using a 200 keV Si beam with a 25 nm FWHM and detection efficiency of 100%, we find the following: (1) the pitch will be limited by the beam spot size on target. Assuming a Gaussian beam profile and an array of PMMA masking holes 25 nm in diameter then we find the minimal pitch for > 95% probability of hitting the correct mask opening is 38 nm. This is based on the overlap of the Gaussian beam profile over the targeted implant site and adjacent sites. (2) To ensure that we implant only one atom per site, it is necessary to implant with on average 0.1 ions per pulse. Under this condition, it will take an average of 12 implant attempts; the cumulative probability of implanting two or more atoms per site is 5.6%. Our fast blanking and chopping system can be run at 400 kHz giving a total time of ~30 μs per implant attempt. Note − our measured drift of 10 nm/minute is negligible for this time period. (3) Using the existing combination of direct write lithography system and CAD navigation, we can implant a 4-inch wafer but this can be extended to larger wafer size. We have demonstrated a stitching resolution of ~50 nm allowing for wafer scale fabrication. In sum, we have the ability to fabricate deterministic single atom devices at 1 per 38x38 nm$^2$ per 30 μs with > 95% confidence using this 'top-down' approach.

The core components of this implant system can be applied to fabrication of devices in different materials. Consider acceptor-based qubits in Si phononic cavities [40]. This proposal requires coupling of single boron atom to a Si crystal cavity. Deterministic coupling is expected to occur when the acceptor is < λ/2 (λ = 385 nm) from the cavity's center. This proposal is compatible with Si based single ion detectors and our demonstrated < 35 nm positioning is sufficient for fabrication. In diamond, we consider cavity QED experiments where single SiV color centers are coupled to a photonic cavity. As in the previous example, deterministic SiV-cavity coupling is expected when the ion is placed < λ/2 (λ is 738 nm) from the center of the cavity, as demonstrated in [41] with an offset of 280 nm. Noting that deterministic coupling of diamond nano-cavities to color centers has been achieved by mechanically placing a cavity over a color center [10], or by fabricating cavities around pre-characterized color centers [41], the advantages of targeted implantation are clear. Our targeted implantation combined with diamond detectors [19] would enable counted single ion implantation into diamond nanostructures. The positioning of the nucleation sites in GaAs [11], and potentially SiGe [42], for growth of quantum dots within ~30 nm of the photonic crystal cavities mode results in a strong quantum dot-cavity coupling [12]. The necessary alignment resolution is approximately the same as the targeting accuracy demonstrated here and the availability of ion detectors in GaAs [20, 43] can allow for counted implantation at the nucleation sites.

We have implemented a complete system for single atom device fabrication via counted ion implantation. We use the SNL nI to position a single atom to within 35 nm of the targeted location and single ion detectors for deterministic device fabrication with a process flow that is CMOS compatible. The approach outlined in this paper is applicable to additional nanoscale devices such as donor-donor (two qubit), electron shuttling, and acceptor based- qubits in Si and can be extended to single atom and nm-scale devices in other material systems such as diamond and III-V compounds. Importantly, these techniques are reliable and compatible with existing fabrication processes.

**Acknowledgements:**

The authors thank Gyorgy Vizkelethy for helpful discussions on ion detection, and J. Dominguez, and B. Silva for support with device fabrication. This work was performed, in part, at the Center for Integrated Nanotechnologies, a U.S. DOE Office of Basic Energy Sciences user facility. Sandia National Laboratories is a multi-program laboratory operated by Sandia Corporation, a Lockheed-Martin Company, for the U. S. Department of Energy under Contract No. DE-AC04-94AL85000.

**Figures:**

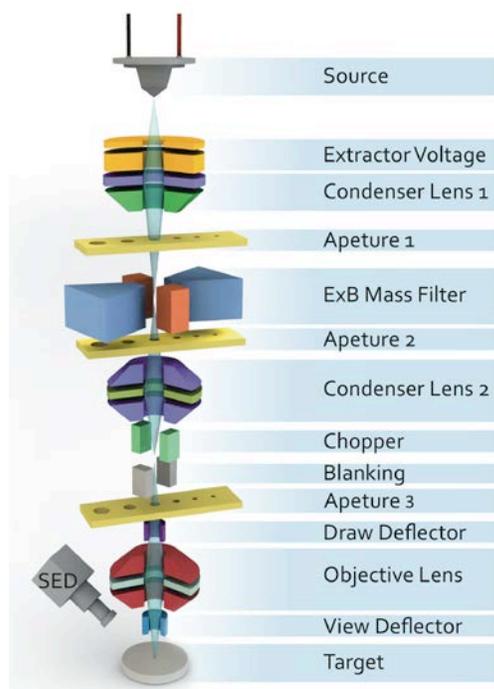

**Figure 1 | Schematic diagram of the nanoImplanter (nI)**. The ion beam is extracted from a liquid metal alloy ion source. The beam crossover at the ExB and at the chopper-blanking regions defines the mass and the beam chopping resolution, respectively. The objective lens focuses the beam on the target plane. The Raith patterning system aligns the beam (rotation, shift, and magnification) with respect to the sample using the draw deflectors and secondary electron detection (SED) images of registration marks.

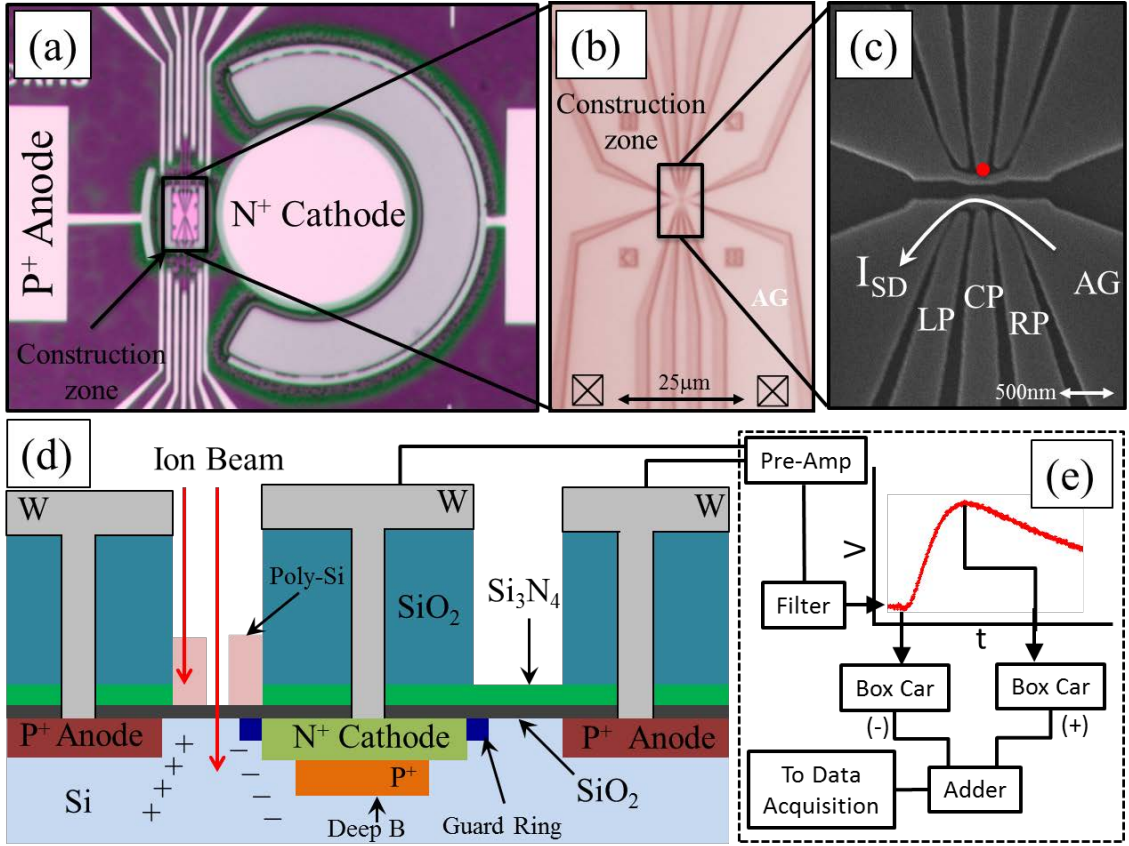

**Figure 2 | Single ion detector with integrated nanostructure.** (a) Top view of detector (concentric circles) with the construction zone patterned in the active detection region. (b) Optical image of nanostructure patterned within the construction zone, accumulation gate (AG) is between source and drain connections (◻). (c) SEM image of the central portion of the nanostructure, light areas are poly-Si gates and darker areas is where the poly-Si has been removed, leaving only 7 nm $SiO_2$ on top of the Si substrate. The red dot shows the targeted donor location. (d) Cross-sectional view of detector (see text for description). Ions that go between poly-Si gates and through the $SiO_2$ layer create e-h pairs in the detector, giving rise to an ion beam induced charge collection (IBIC) signal. (e) Measurement circuit consisting of a pre-amplifier, two box-cars, and an adder that outputs the amplitude of the response.

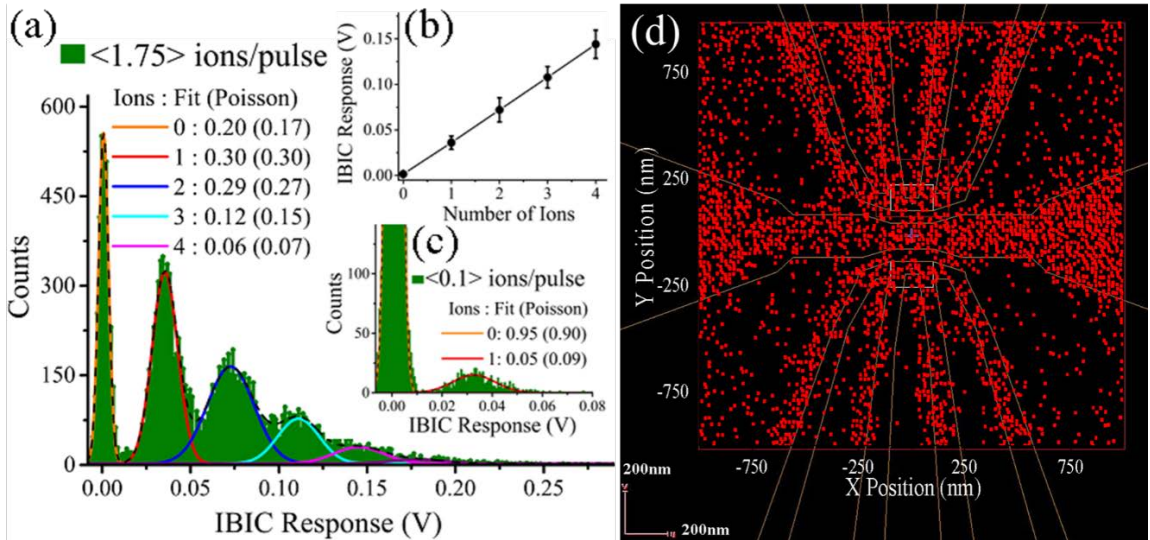

**Figure 3 | Quantization ion detection and ion beam targeting**. (a) IBIC response to 200 keV Si beam with an average of 1.75 ions per pulse. The probability of a certain number of ions being implanted and each of these agree to ± 3% of the probability predicted by Poisson statistics. (b) Calibration of detector response; black dots show the detector response for 1, 2, 3, and 4 ions per pulse. The response is ~0.037 V per 200 keV Si ion. (c) Response to an average of 0.1 ions per pulse. (d) An overlay of a targeted 2 μm by 2 μm array of implant locations that show IBIC response and the nanostructure CAD as background. The red square outlines the location of the array exposed as defined in the patterning software, see text for details.

## Tables:

**Table 1|** Signal-to-noise ratio, detection error rates and detection efficiency (where available) are shown for the different combinations of ion species, implantation energy, and over-layer thickness.

| Ion | Energy (keV) | SiO$_2$ Thickness | Range*(± Straggle) | e-h Pairs* | SNR | Error Rate | Det. Eff. |
|---|---|---|---|---|---|---|---|
| Si | 200 | 7 nm | 273 (± 76) nm | 39k | 21.2 | <<1% | 100% |
| Sb | 120 | 35 nm | 25 (± 17.5) nm | 8.5k | 5.2 | -- | -- |
| Sb | 50 | 7 nm | 25 (± 9) nm | 5.0k | 4.4 | -- | -- |
| Sb | 20 | 7 nm | 11 (± 5) nm | 1.4k | 2.5 | 15% | 87.6% |

The error rate of detection is defined as 1-P(TP) = P(FN), where P(TP) and P(FN) are the probability of true positive and false negative detection, respectively and using the standard deviation of the noise peak as a reference [39]. * Range and e-h pairs calculated from the SiO$_2$/Si interface into the Si substrate.